\documentclass[pra,aps,preprint,showpacs]{revtex4-1}
\usepackage{amsmath}
\usepackage{amssymb}
\usepackage{graphicx}
\DeclareGraphicsExtensions{.eps}

\newcommand{\mP}{{\mathcal P}}
\newcommand{\mT}{{\mathcal T}}
\newcommand{\mC}{{\mathcal C}}

\newcommand{\br}{{\bf r}}

\newcommand{\beqa}{\begin{eqnarray}}
\newcommand{\eeqa}{\end{eqnarray}}

\begin{document}
\title{Mapping between Hamiltonians with attractive and repulsive potentials on a lattice} 
\author{Yogesh N. Joglekar}
\affiliation{Department of Physics, 
Indiana University Purdue University Indianapolis (IUPUI), 
Indianapolis, Indiana 46202, USA}
\date{\today}
\begin{abstract}
Through a simple and exact analytical derivation, we show that for a particle on a lattice, there is 
a one-to-one correspondence between the spectra in the presence of an attractive potential $\hat{V}$ and its repulsive counterpart $-\hat{V}$. For a Hermitian potential, this result implies that the number of localized states is the same in both, attractive and repulsive, cases although these states occur above (below) the band-continnum for the repulsive (attractive) case.  For a $\mP\mT$-symmetric potential 
that is odd under parity, our result implies that in the $\mP\mT$-unbroken phase, the energy eigenvalues are symmetric around zero, and that the corresponding eigenfunctions are closely related to each other. 
\end{abstract}
\maketitle

\noindent{\it Introduction:} The energy spectrum of a quantum particle in an attractive potential 
$V(\br)$, in general, consists of discrete eigenvalues for which the eigenfunctions are localized in real space, and continuum eigenvalues with non square-integrable eigenfunctions. The energy spectrum for the corresponding repulsive potential $-V(\br)$ has only continuum eigenvalues~\cite{qmbasics,caveat}. This situation changes dramatically when the particle is confined to a lattice or, equivalently, is exposed to a periodic potential. Indeed, repulsively bound two-atom states have been explored in detail since their experimental discovery in optical lattices~\cite{winkler,fallani} and continue to be a source of ongoing work~\cite{mahajan} in the context of the Bose-Hubbard model~\cite{valiente,wang}. 
We note that in the Bose-Hubbard model, the interaction between the two atoms is short-ranged and is tuned via the Feschback resonance~\cite{winkler}. However, to our knowledge, the properties of single-particle states localized in the vicinity of a generic repulsive potential (defined below) have not been studied. In another area, localized states in parity + time-reversal ($\mP\mT$) symmetric one-dimensional lattice models, too, have been explored in recent years. These explorations have 
focused on the $\mP\mT$-symmetry breaking in the presence of attractive (real) on-site potentials with random $\mP\mT$-symmetric complex parts~\cite{bendix}. 

In this note, through a simple but exact derivation, we show that for a single particle on a lattice, there is a one-to-one correspondence between its energy spectrum in the presence of an attractive potential and the repulsive counterpart, and that the corresponding eigenfunctions have identical probability distributions. For $\mP\mT$-symmetric potentials that are odd under parity (and hence time-reversal), we show that if the $\mP\mT$-symmetry is unbroken, the energy spectrum must be symmetric around zero. 

\noindent{\it One-dimensional Model:} 
Let us start with the Hamiltonian for a particle on a one-dimensional lattice with only nearest-neighbor hopping energy $J>0$, 
\begin{equation}
\label{eq:h0}
\hat{H}_0=-J\sum_i\left( c^{\dagger}_{i}c_{i+1} + c^{\dagger}_{i+1}c_i\right)
\end{equation}
where $c^{\dagger}_i$ and $c_i$ are creation and annihilation operators at site $i$ respectively. The external potential is given by $\hat{V}=\sum_j V_j c^{\dagger}_j c_j$. We define the potential to be attractive provided $\sum_j V_j<0$ and repulsive if is positive. Let $|\psi_\alpha\rangle=\sum_j f_{\alpha,j}|j\rangle$ be an eigenstate of the Hamiltonian $\hat{H}_{+}=\hat{H}_0+\hat{V}$ with energy $E_\alpha$ where $|j\rangle$ denotes a single-particle state localized at site $j$. The coefficients $f_{\alpha,j}$ obey the recursion relation 
\begin{equation}
\label{eq:recursion}
-J\left[f_{\alpha,j+1}+f_{\alpha,j-1}\right]+ V_j f_{\alpha,j}=E_\alpha f_{\alpha,j}.
\end{equation}

We now consider the staggered wavefunction $|\phi_\alpha\rangle=\sum_j f_{\alpha,j}(-1)^{j}|j\rangle$. 
Using Eq.(\ref{eq:recursion}) it is straightforward to show that the staggered wavefunction satisfies 
the following equation 
\begin{equation}
\hat{H}_0 |\phi_\alpha\rangle=\left(- E_\alpha +\hat {V}\right)|\phi_\alpha\rangle.
\end{equation}
Thus, it is an eigenfunction of the conjugate Hamiltonian $\hat{H}_{-}=\hat{H}_0-\hat{V}$ 
with eigenvalue $-E_\alpha$. When $\hat{V}=0$, the energy spectrum is given by $\epsilon_k=-2J \cos(ka)$ and represents the well-known continuum band from $-2J$ to $2J$ where $a$ is the lattice spacing. In this trivial case, indeed the eigenfunction $|\psi_k\rangle=\sum_j\sin(k j)|j\rangle$ and its staggered counterpart $|\phi_k\rangle=\sum_j\sin\left[(\pi-k)j\right]|j\rangle$ have energies $\pm\epsilon_k$ respectively.  

\begin{figure}[h!]
\begin{center}
\includegraphics[angle=0,width=17cm]{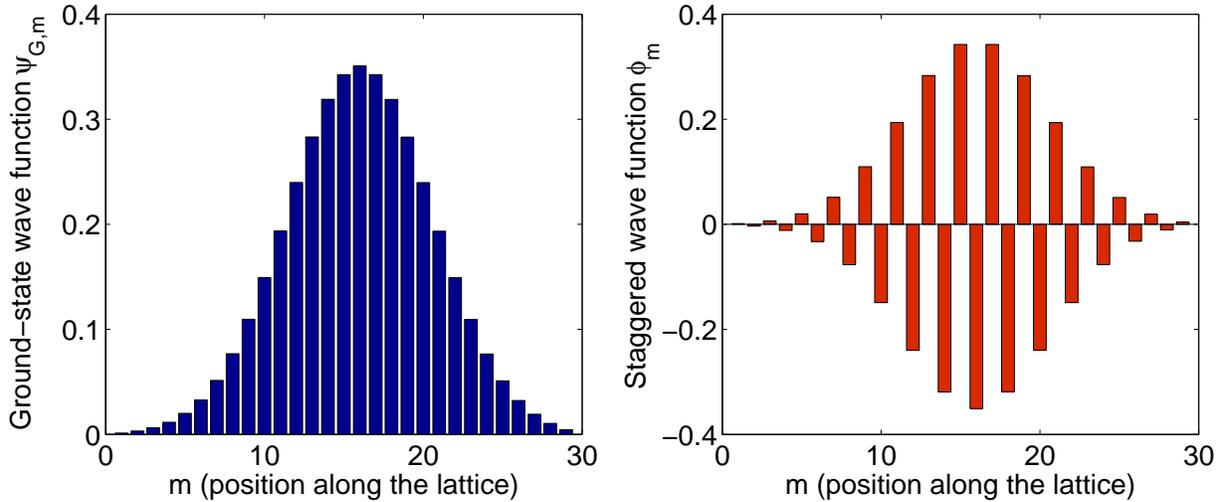}
\caption{\label{fig:gauss}
(color online) (a) The left panel shows the dimensionless ground-state wavefunction $\psi_{G,m}$ for an {\it attractive quadratic potential} $V_m=\Lambda(m-1)(N-m)/N_0^2$ where $N=29=(2N_0+1)$ is the lattice size and $\Lambda/J=-0.5$. As expected for a quadratic potential ground-state, $\psi_{G,m}$ is a Gaussian with 
width $x_0=a (N_0^2t/|\Lambda|)^{1/4}\sim 4.61$. (b) The right panel shows the dimensionless highest-energy state wavefunction $\phi_m$ for its {\it repulsive counterpart} with $\Lambda/J=+0.5$. We see that the $\phi_m$ is indeed the staggered version of the ground-state wavefunction $\phi_{G,m}$.}
\end{center}
\end{figure}

Our result shows that if an attractive external potential $\hat{V}$ has $n$ bound states below its continuum  with energies $E_m$ ($m=1,\ldots, n$), then the corresponding repulsive potential $-\hat{V}$ must have an equal number of bound states above its continuum with energies $-E_m$. Since 
the staggered wavefunction $|\phi_\alpha\rangle$ varies over the lattice length-scale $a$, it is 
energetically expensive and ill-defined in the continuum limit $a\rightarrow 0$. Physically, in the 
continuum limit, the absence of lattice-site scattering centers makes it impossible for a particle to 
localize near the repulsive potential. However, on a lattice, the probability distributions for the two states - a localized bound state $|\psi_\alpha\rangle$ with energy $E_\alpha\leq -2J$ in an attractive potential and the localized bound state $|\phi_\alpha\rangle$ with energy $-E_\alpha\geq +2J$ in the repulsive potential - are identical.  As a concrete example, we numerically obtain the spectrum for a lattice with $N=29$ sites and a quadratic potential that vanishes at the ends, $V_m=\Lambda (m-1)(N-m)/N_0^2$, 
where $m=1,\ldots,N$, $N_0=(N+1)/2$ is the center of the lattice and $V_{N_0}=\Lambda$. Figure~\ref{fig:gauss} shows the ground state wavefunction $\psi_{Gm}$ for the attractive case, $\Lambda/J=-0.5$, (left panel) along with the highest-energy state wavefunction $\phi_m$ for the repulsive case, $\Lambda/J=+0.5$ (right panel). It is clear that the two wavefunctions are related by $\phi_m=(-1)^{m+1}\psi_{G,m}$. 

\noindent {\it Two-particle Case:} We can generalize this result in a straightforward manner to treat 
interparticle interaction $\hat{U}=\sum_{ij} U_{i-j} \hat{n}_i\hat{n}_j$ where the on-site number operator is given by $\hat{n}_i=c^\dagger_i c_i$. In the two-particle sector, the recursion relation satisfied by the relative-coordinate wavefunction  is given by~\cite{valiente,wang}
\begin{equation}
\label{eq:2particle}
-J_K\left[\psi^{K}_{\alpha,m+1}+\psi^{K}_{\alpha,m-1}\right]+U(r_m)\psi^{K}_{\alpha,m}= E^{K}_\alpha\psi^{K}_{\alpha,m}.
\end{equation}
Here $-\pi/a\leq K\leq \pi/a$ is the lattice momentum associated with the center-of-mass of the 
two particles, $J_K=J\cos(Ka)$ is the effective hopping energy, $r_m=am=a(i-j)$ is the distance between the two particles on the lattice located at sites $i$ and $j$, and $U(r_m)$ is the real-space interaction between the two particles. Note that for a non-local interparticle interaction $U(r_m)$, multiple bound-state $\psi^{K}_\alpha$ solutions are generic, although, in the context of the Bose-Hubbard model, only one~\cite{winkler} or two~\cite{valiente} have been discussed. If $\psi^K_\alpha$ is an eigenfunction of the Hamiltonian $\hat{H}_0+\hat{U}$ with energy $E^K_\alpha$,  Eq.~\ref{eq:2particle} implies that the staggered wavefunction $\phi^K_\alpha$ defined by $\phi^K_\alpha(r_m)=(-1)^m\psi^K_\alpha(r_m)$ is an eigenfunction of the conjugate Hamiltonian $\hat{H}_0-\hat{U}$ with energy $-E^K_\alpha$.

Two-particle bound states in the presence of on-site and nearest-neighbor repulsive density-density interactions on a lattice have been extensively investigated~\cite{winkler,mahajan,valiente}. Our derivation shows that they are a generic feature of any density-density interaction on a lattice, and this result is true for square lattices in higher dimensions. Note that the quantum statistics of the particles only constrains the relative wavefunction $\psi^K_\alpha(r_m)$ to be odd (spinless fermions) or even (bosons or spin-singlet fermions) under parity; however, it does not affect the one-to-one 
correspondence between the spectra for the two Hamiltonians $\hat{H}_0\pm\hat{U}$. Thus, two-atom 
bound-states with attractive and repulsive interactions in optical lattices (bosons)~\cite{winkler}, the 
donor and acceptor impurity levels in semiconductors (fermions)~\cite{cardona}, as well as the 
localized phonon modes (collective bosonic excitation)~\cite{chen,mizuno} around a soft or 
stiff impurity can all be thought of as manifestations of the correspondence between 
spectra for $\hat{H}_{+}$ and $H_{-}$.

\noindent $\mP\mT$ {\it Symmetric Potential:} The mapping between the two Hamiltonians $\hat{H}_{+}$ and $\hat{H}_{-}$ is valid independent of the properties of the potential $\hat{V}$ including its Hermiticity; the on-site potential elements $V_j$ may be complex. However, for a $\mP\mT$-symmetric potential that is odd under parity (and hence, time reversal), $V^{*}_j=-V_j=V_{-j}$, it follows that $\hat{H}^*_{+}=\hat{H}_{-}$ where * denotes complex conjugation. Therefore, it follows from $\hat{H}_{+}|\psi_\alpha\rangle=E_\alpha |\psi_\alpha\rangle$ that the wavefunction $|\psi^*_\alpha\rangle=\sum_j f^*_{\alpha,j}|j\rangle$ is an eigenstate of the conjugate Hamiltonian $\hat{H}_{-}$ with eigenvalue $+E^*_\alpha$. In the continuum limit, it has been shown that a wide class of such potentials, including $V(x)=ix^3$ and $V(x)= i\sin^{2n+1}(x)$ have purely real energy spectra~\cite{bender,bender2}. If the $\mP\mT$-symmetry is unbroken, $E^*_\alpha=E_\alpha$, then it follows that $\hat{H}_{-}|\phi_\alpha\rangle= -E_\alpha|\phi_\alpha\rangle$ and $\hat{H}_{-}|\psi^{*}_\alpha\rangle=+E_\alpha|\psi^{*}_\alpha\rangle$. 

This explicit construction of wavefunctions with equal and opposite energies implies that for any arbitrary $\mP\mT$-symmetric potential that is odd under parity, if the $\mP\mT$ symmetry is not broken, the energy spectrum must be symmetric around zero. It also shows that the corresponding wavefunctions in the two cases have components that are simply related: $[+E_\alpha, f^{*}_{\alpha,j}]\leftrightarrow [-E_\alpha,f_{\alpha,j}(-1)^j]$. As an example, we consider the simplest ``finite lattice" with 2 points. (Our result is equally applicable to a finite lattice.) The Hamiltonian in this case is given by $\hat{H}_{-}=-J\hat{\sigma}_x+i\gamma\hat{\sigma}_z$ where $(\sigma_x,\sigma_z)$ are the Pauli matrices in the site-index space~\cite{bender3} and a real $\gamma$ ensures that the potential is odd under parity as well as time-reversal.  The eigenvalues in this case are given by $E_{\pm}=\pm\sqrt{J^2-\gamma^2}$. Thus the $\mP\mT$-symmetry in this case is not broken as long as $\gamma\leq J$. The corresponding (unnormalized) eigenfunctions~\cite{innerproduct} are given by~\cite{bender3}
\begin{equation}
\label{eq:soln}
|\pm\rangle = \left(
\begin{array}{c}
1 \\
\pm e^{\mp i\theta}\\
\end{array}\right)
\end{equation}
where $\theta=\arctan(\gamma/\sqrt{J^2-\gamma^2})$ is real when $\gamma\leq J$. Therefore, in the 
$\mP\mT$-unbroken phase, the eigenvectors for positive and negative energies indeed are related by 
$f_{-,j}=(-1)^j f^{*}_{+,j}$ where $j=0,1$. 

\noindent{\it Conclusion:} Our result, through a one-to-one mapping between attractive and repulsive 
potentials on a lattice, shows that localized states in repulsive potentials are ubiquitous. These states 
can be explored via local measurements. In contrast to the bound-states with energies below the 
continuum band, these localized states with energies above the continuum band will decay into the 
continnum states. They may thus provide a useful spectroscopic tool in optical lattices as well as 
engineered electronic materials with a small bandwidth. 





\begin{thebibliography}{99}
\bibitem{innerproduct} We consider unnormalized eigenvectors to avoid the subtleties, associated with 
the $\mC\mP\mT$ inner-product~\cite{bender3}, that are irrelevant for our result.
\end{thebibliography}
\end{document}